\def\MCFM{{\tt MCFM}}
\def\PYTHIA{{\tt PYTHIA}}
\def\HERWIG{{\tt HERWIG}}
\def\POWHEGBOX{{\tt POWHEG BOX}}
\def\MadaMCNLO{{\tt MaGraph5\_aMC@NLO}}
\def\PowHel{{\tt PowHel}}
\def\SHERPA{{\tt SHERPA}}
\def\OpenLoops{{\tt OpenLoops}}

\def\nlox{{\tt NLOX}}

\documentclass[a4paper]{PoS}

\usepackage{url}
\usepackage{float}
\usepackage{subfigure}

\title{Precision studies of vector-boson production with heavy
  quarks at the LHC: the case of $Z+b$ jet.}

\ShortTitle{Vector boson production with heavy quarks}

\author{D.~Figueroa,$^a$ S.~Honeywell,$^a$ S.~Quackenbush,$^a$ 
\speaker{L.~Reina},$^a$ C.~Reuschle$^a$ and D.~Wackeroth$^b$\\
\llap{$^a$}Physics Department, Florida State University,\\ Tallahassee, FL 32306-4350, U.S.A.\\
\llap{$^b$}Department of Physics, SUNY at Buffalo,\\ Buffalo, NY 14260-1500, U.S.A.\\
E-mail: \email{daf14f@my.fsu.edu}, \email{sjh07@hep.fsu.edu}, \email{squackenbush@hep.fsu.edu}, \email{reina@hep.fsu.edu}, \email{creuschle@hep.fsu.edu}, \email{dow@ubpheno.physics.buffalo.edu}}

\abstract{After a brief introduction to the state of the art of
  theoretical predictions for electroweak vector-boson production with
  both top and bottom quarks at the LHC, we review the case of 
  $Z$-boson production with $b$ jets, and discuss the impact of
  NLO QCD+EW corrections and finite $b$-mass effects on the 
 theoretical prediction for $Z+b$ jet production.}

\FullConference{Loops and Legs in Quantum Field Theory (LL2018)\\
		29 April 2018 - 04 May 2018\\
		St. Goar, Germany}

\begin{document}

\section{Overview and Outlook}

The production of electroweak (EW) vector bosons ($V=W^\pm,Z$)
with heavy quarks ($b$ and $t$) is particularly relevant to the
physics of the Large Hadron Collider (LHC) for several reasons.

First of all, the associated production of $W$ or $Z$ bosons with
heavy quarks constitutes an important background for Higgs physics and
searches of new physics beyond the Standard Model (SM). For example,
$V+t\bar{t}$ with $V$ decaying into leptons is the dominant background
for measuring the production of a SM Higgs boson in the $t\bar{t}H$
channel followed by $H\rightarrow W^+W^-$(with subsequent decay to
$l^+\nu l^-\bar{\nu}$ or $l\nu q\bar{q}^\prime$), $H\rightarrow ZZ$ (with subsequent decay to
$\nu\bar{\nu}q \bar{q}$ or $l^+l^- q\bar{q}$), or
$H\rightarrow \tau^+\tau^-$. On the other hand, $W$ and $Z$ production
with $b$ jets are the dominant backgrounds in the measurement of
associated Higgs-boson production with either $W$ or $Z$, when the
Higgs-boson decays into $b$ quarks ($H\rightarrow b\bar{b}$). Hence
the accurate theoretical prediction of both $V+t\bar{t}$ and $V+b$
jets is a crucial element in testing the Higgs-boson couplings to
third-generation quarks.

At the same time, both $V$ production with top and bottom quarks have
their own intrinsic value in testing important aspects of weak and
strong interactions. Measuring $V+t\bar{t}$ production provides a
direct test of anomalies in top-quark electroweak couplings, while
$W+t$ represents one of the components of single-top production, whose
measurement could help constraining the $Wtb$ coupling from single-top
measurements. If $V+t\bar{t}$ and $V+t$ processes are specific domain
of the LHC, given the massiveness of the particles produced, the
analogous $V+b$-jet processes have been measured already at the
Tevatron and sometimes left us with puzzling results that are now
being revisited by the LHC. From the first round of LHC measurement we
can see emerging a much more consistent picture, and we can start
envisaging the possibility of using these processes as a testing
ground to understand more complicated signatures involving $b$ jets
(e.g. crucial background processes such as $t\bar{t}+b$ jets).  Last
but not least, increasing experimental precision and theoretical
accuracy in the prediction of $V+b$-jet processes can provide a direct
measurement of the $b$-quark parton distribution function (PDF), whose
direct knowledge could constrain the theoretical parameterizations
currently used in PDF fits.

Theoretical progress in providing more and more accurate predictions
for both $V+t\bar{t}$ and $V+b$ jets has been steady over the last few
years. The Next-to-Leading Order (NLO) QCD corrections to $Z+t\bar{t}$
and $W^\pm+t\bar{t}$ production at the LHC have been calculated
in~\cite{Lazopoulos:2008de,Kardos:2011na} and \cite{Campbell:2012dh}
respectively, and interfaced with parton-shower
in~\cite{Garzelli:2011is,Garzelli:2012bn} in the \PowHel\, framework.
The complete NLO QCD and electroweak (EW) cross sections for
$t\bar{t}+V$ hadroproduction have more recently been
presented in~\cite{Frixione:2015zaa}, and successfully compared with
results from a calculation based on
\OpenLoops+\SHERPA~\cite{Kallweit:2014xda} in the context of the Higgs
Cross Section Working Group~\cite{deFlorian:2016spz}.  In the same
context~\cite{deFlorian:2016spz}, differential cross sections from
existing fixed-order NLO QCD calculations of $V+t\bar{t}$ have been
successfully cross checked. The theoretical uncertainty of the NLO
QCD+EW $V+t\bar{t}$ total cross sections, due to both renormalization
and factorization scale dependence, PDF, and $\alpha_s$, is estimated
to be around 15-20\%, much improved if compared to the corresponding
tree-level results. On the other hand, the cross sections for $W+t$
production has been calculated at NLO~\cite{Zhu:2001hw} and the
improved cross section with resummation of next-to-next-to leading
logarithms (NNLL) has been presented in~\cite{Kidonakis:2010ux}, where
the theoretical uncertainty is found to be  around a few percents.

The case of $V+b$-jet hadroproduction has a more intriguing
theoretical structure, due to both its intrinsic dependence on largely
different scales ($m_b$, $M_V$, and the characteristics energy of LHC
collisions), which introduces the possibility of large logarithmic
contributions in particular kinematic regions and energy regimes and
prompt to their resummation, and to the possibility of also involving
$b$-initiated processes.  On top of refining the theoretical
predictions by adding higher-orders of quantum corrections, theorists
have also been debating the best approach to the calculation of the
$V+b$-jet cross sections, with the aim of controlling and improving
the stability of perturbative predictions. With this respect, at the
NLO of QCD+EW corrections, the calculation of the hadronic production
of $V+2b$ jets can only proceed in strict analogy with
$V+t\bar{t}$. On the other hand, the case of $V+1b$ jet can be
calculated assuming or not assuming an initial-state $b$ density,
i.e. using a 5-flavor scheme (5FS) or a 4-flavor scheme (4FS). In the
first case (5FS) the definition of a $b$-quark PDF removes from the
hard cross-sections some of the large logarithmic corrections due to
initial-state radiation and resums them via renormalization-group
scale evolution. In the second case (4FS) a more realistic description
of hard final-state radiation can be achieved.  Being the 4FS and 5FS
just two different ways of reordering the perturbative expansion of
the cross section, their predictions agree better and better the
higher the perturbative order reached in a calculation. More extensive
discussions of this issue has been presented in the
literature~\cite{Maltoni:2012pa,Forte:2015hba,Cordero:2015sba,Lim:2016wjo,
  Forte:2016sja,Bonvini:2015pxa,Bonvini:2016fgf,Krauss:2016orf}.

For $V+b$-jet processes, most calculations so far have focused on
higher-order QCD corrections and interfacing fixed-order NLO QCD
calculations with parton-shower event generators. In particular, NLO
QCD corrections have been calculated for $W+2j$ with at least one $b$
jet~\cite{Campbell:2006cu,Campbell:2008hh,Caola:2011pz}, $W+2b$
jets~\cite{FebresCordero:2006sj,Cordero:2009kv,Badger:2010mg}, $W+2b$
jets plus a light jet~\cite{Luisoni:2015mpa,Reina:2011mb}, $Z+1b$
jet~\cite{Campbell:2003dd}, $Z+2j$ with at least one $b$
jet~\cite{Campbell:2005zv}, $Z+2b$
jets~\cite{FebresCordero:2008ci,Cordero:2009kv}, where, depending on the
process, either the 4FS or the 5FS have been used. The theoretical
uncertainty of NLO QCD calculations, in particular for $W+b$ jets
remains quite large, of the order of 20-25\% or more, due to the
opening of new large tree-level contributions among the real radiation
processes.  Recently, a very interesting study~\cite{Anger:2017glm}
that has pushed the NLO QCD calculation of $W+2b$ jets plus $n$ light
jets up to $n=3$ has been able to confirm that most residual scale
uncertainty can indeed be cured by including higher-order corrections
beyond NLO QCD. Although technically very challenging, this could be
within reach in the close future. At the same time, realistic studies
interfacing NLO QCD fixed-order calculations with parton-shower
(\PYTHIA~\cite{Sjostrand:2006za,Sjostrand:2007gs} and
\HERWIG~\cite{Marchesini:1991ch,Corcella:2000bw}) have been presented
in the \POWHEGBOX~\cite{Alioli:2010xd} framework for $W+2b$
jets~\cite{Oleari:2011ey} as well as $W+2b$ jets plus one light
jet~\cite{Luisoni:2015mpa}, and in the
\MadaMCNLO~\cite{Alwall:2014hca} framework for $W/Z+2b$
jets~\cite{Frederix:2011qg}. A separate study using
\SHERPA+\OpenLoops\, has appeared in~\cite{Krauss:2016orf}.  All these
studies represent the state of the art of theoretical predictions that
have been and are compared with both Tevatron and LHC results
(see~\cite{Cordero:2015sba} and~\cite{Badger:2016bpw} for a detailed discussion of these
comparisons up to results from Run I of the LHC).  Most of the
fixed-order total and differential rates for $V+t\bar{t}$ and $V+b$
jets are available through \MCFM~\cite{mcfm8}.

Quite recently, combined NLO+EW correction to $Z$ plus one $b$-jet
hadroproduction have been presented in~\cite{Figueroa:2018chn}, with
attention to the effect of considering a massive $b$ quark both in the
initial and final state. Results from this study will be presented in
Section~\ref{sec:zb-massive-nloqcd+ew}. It is crucial to notice that,
in spite of their small size, controlling these effects is
important. EW and $m_b$ effects can affect kinematic distributions in
complementary regions, relevant to different measurements. Also, the
case of initial-state massive partons~\cite{Collins:1998rz} deserves
particular attention in implementing a consistent matching between NLO
fixed-order 5FS calculations and parton showers (see
also~\cite{Krauss:2017wmx}).

\section{The case of $Z$ plus one $b$ jet: NLO QCD and EW corrections
  for a massive $b$}
\label{sec:zb-massive-nloqcd+ew}

The combined NLO QCD+EW total and differential cross sections for
$Z+1b$ jet have been calculated in~\cite{Figueroa:2018chn} considering
five flavors and the $b$ quark massive in both initial- and
final-state processes, i.e. in the so called \textit{massive} 5FS (or
m5FS). 

Results for the LHC with center of mass energy 13 TeV have been
presented in~\cite{Figueroa:2018chn}. They have been obtained using
the \nlox\, one-loop provider~\cite{NLOX} for one-loop virtual QCD and
EW corrections, and several in-house codes implementing both
phase-space slicing and dipole subtraction for the real
corrections. In order to take into account all $b$-mass effects, both
the initial-state kinematics, the phase-space integration, the UV
counterterms, and the PDF subtraction terms have been consistently
modified. Full details can be found in~\cite{Figueroa:2018chn}.

\begin{figure}
\centering 
\begin{tabular}{cc}
\includegraphics[scale=0.45]{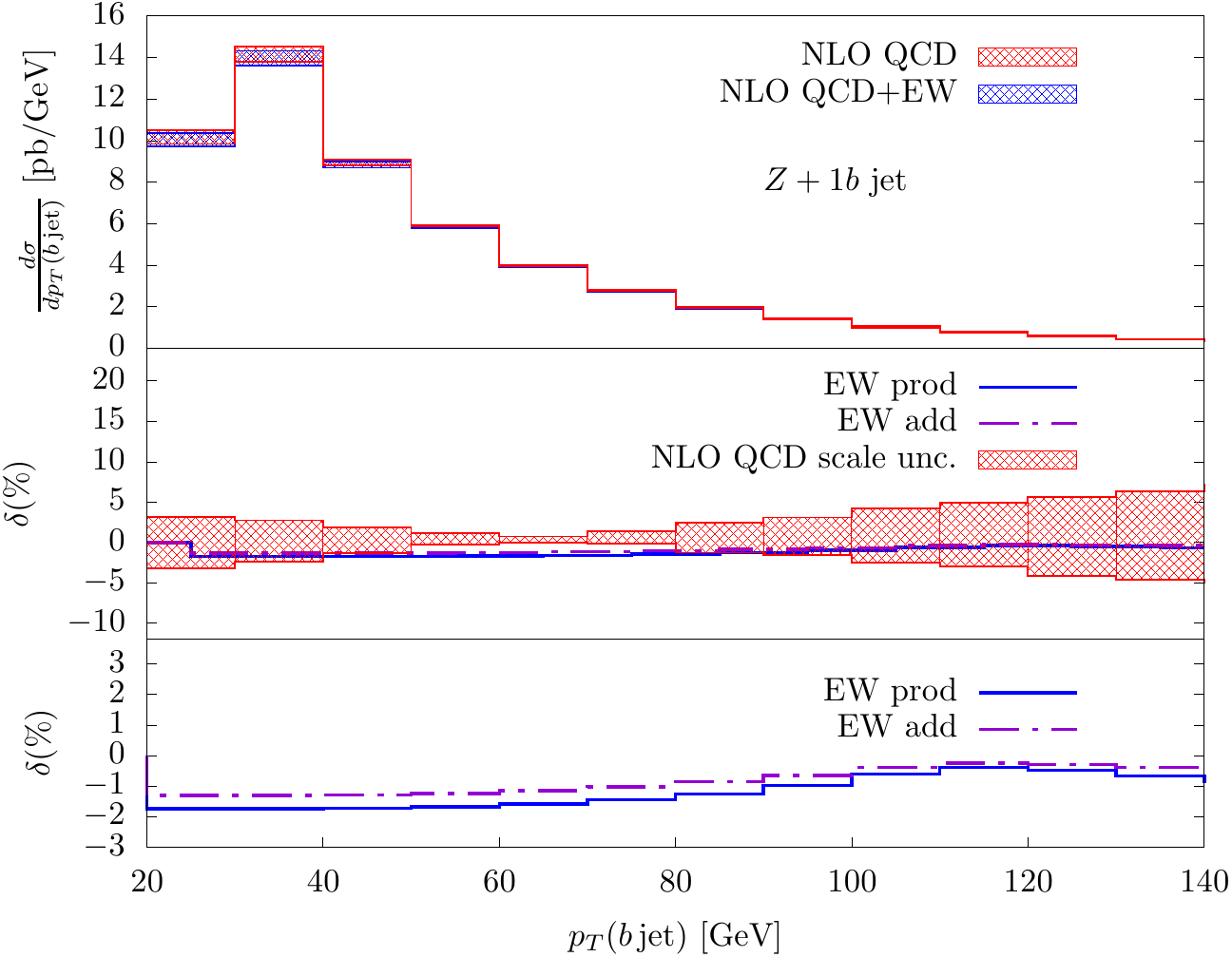} &
\includegraphics[scale=0.45]{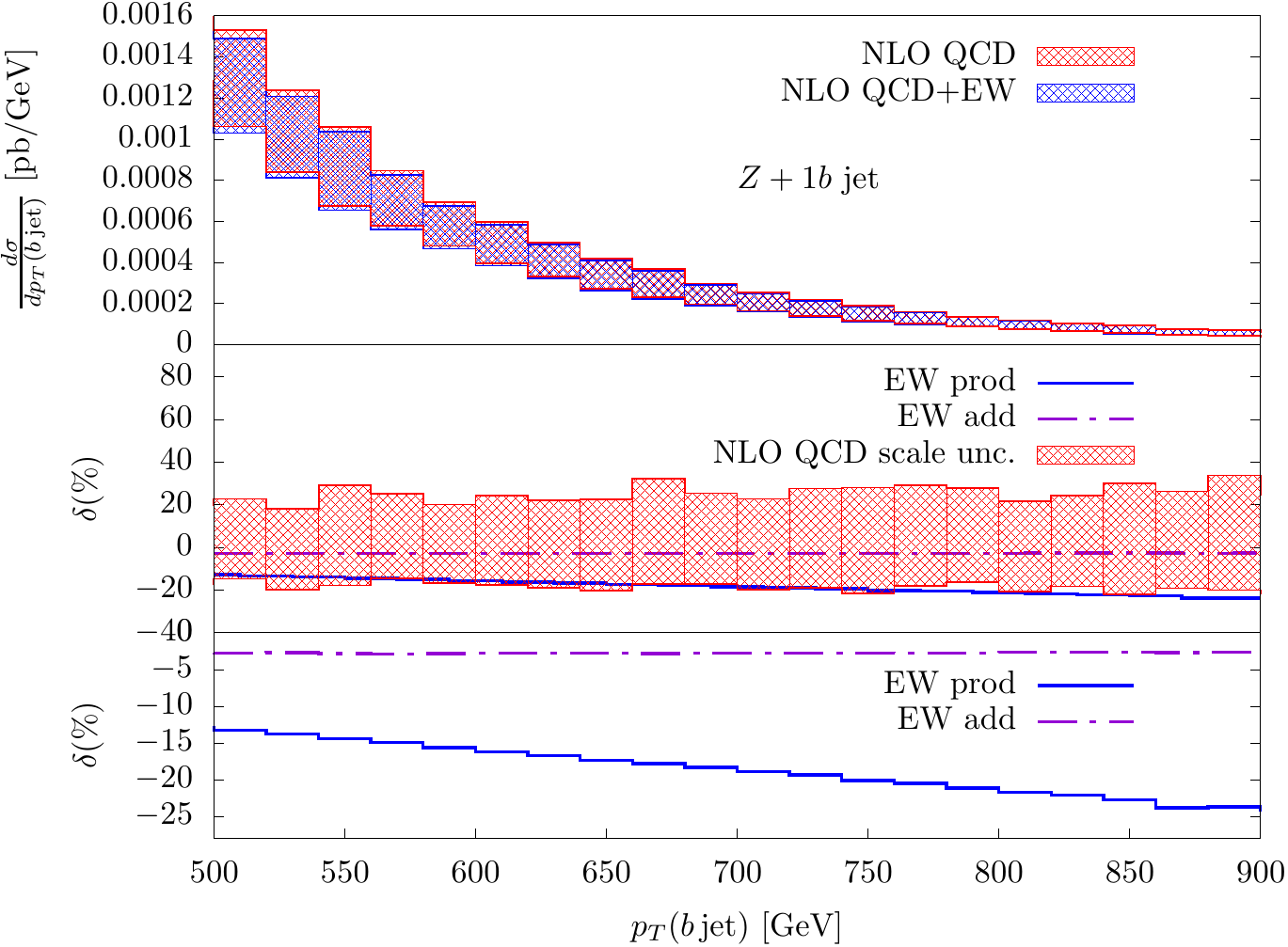}\\
\includegraphics[scale=0.45]{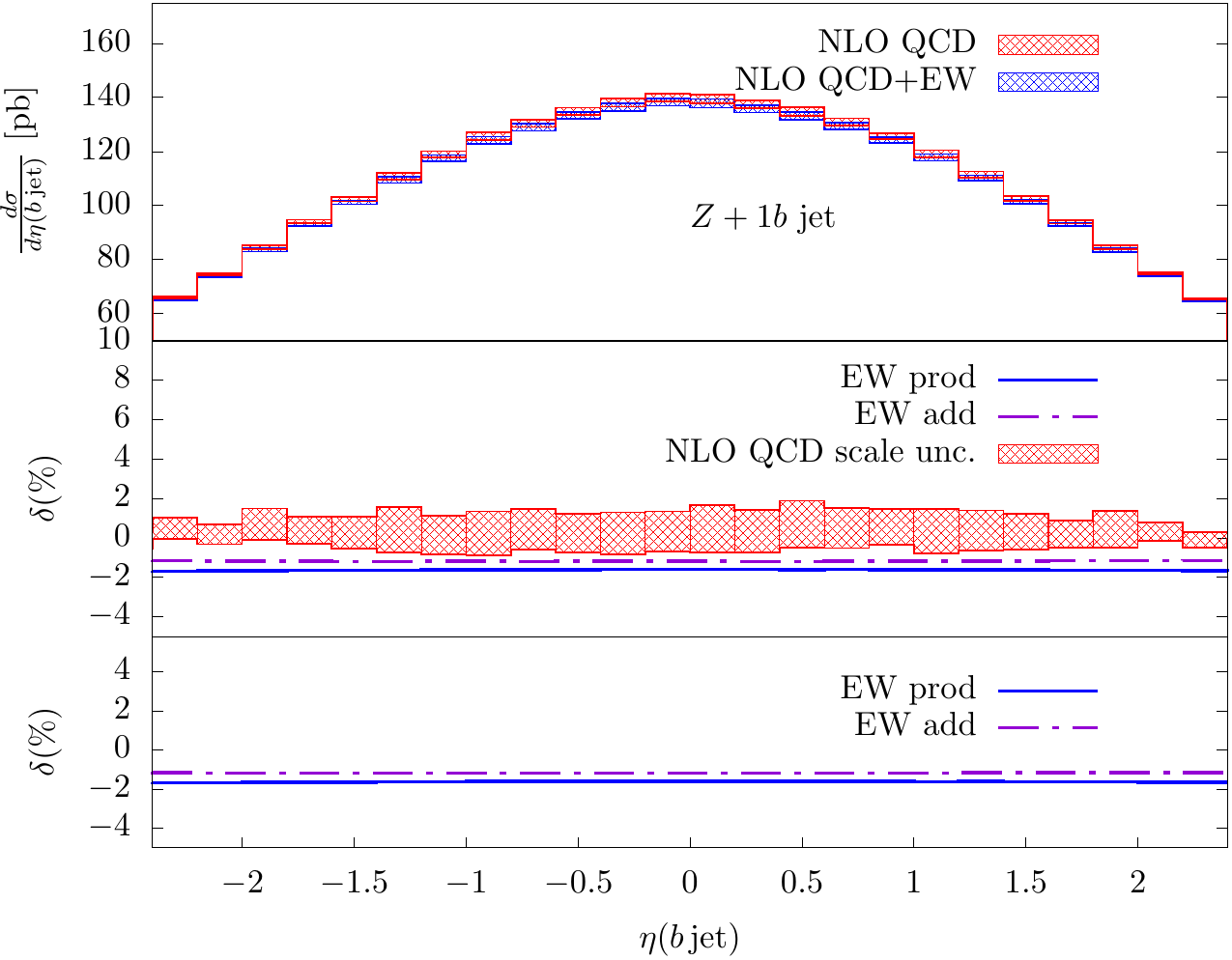} &
\includegraphics[scale=0.45]{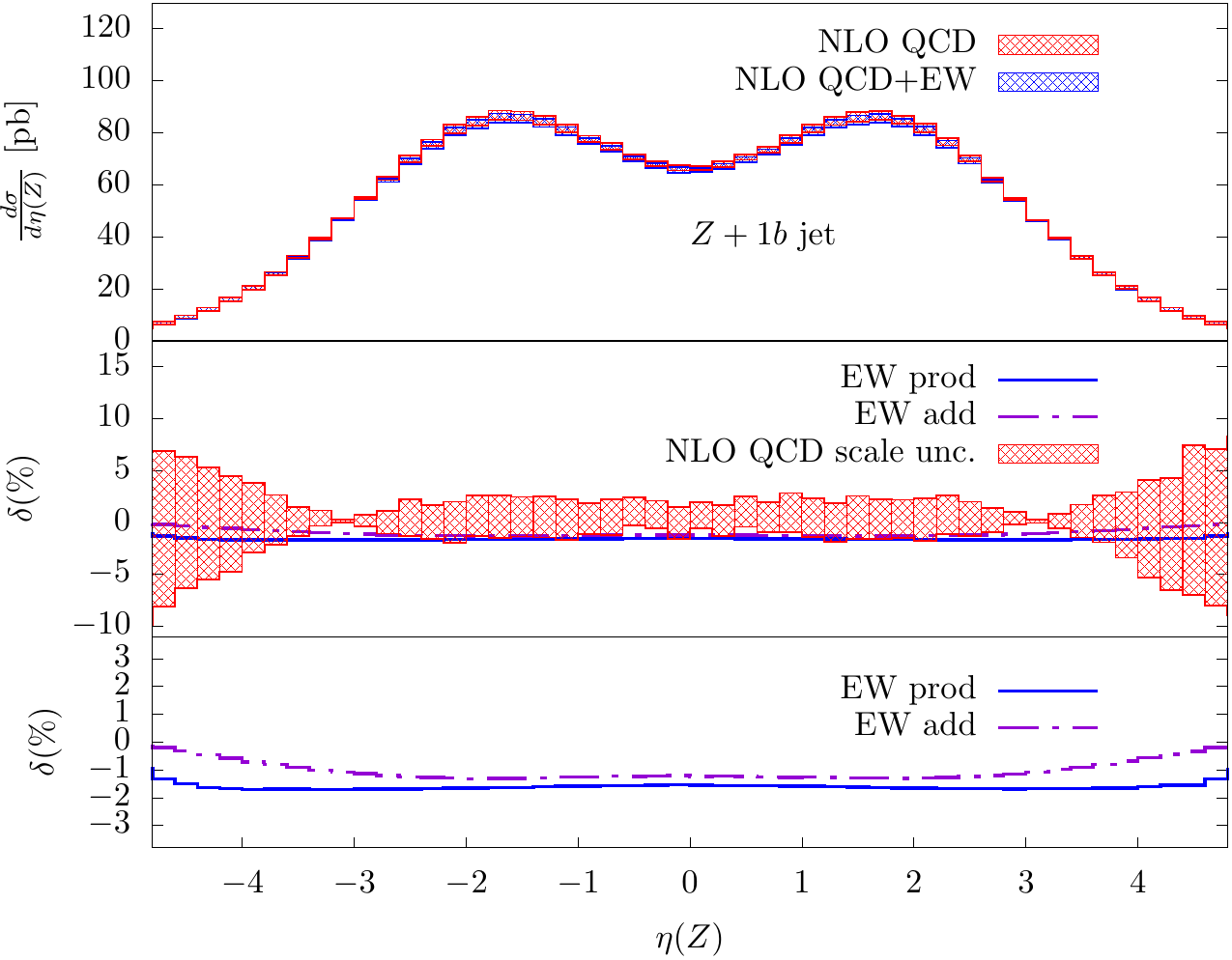}
\end{tabular}
\caption{Differential distributions for the $b$-jet transverse
  momentum, and for the pseudorapidity of both the $b$ jet and $Z$ boson.
The lower windows in each plot show the relative EW $O(\alpha)$ corrections differential
distributions, $\delta_{\mbox{\tiny EW}}^{\mbox{\tiny prod}}$ and
$\delta_{\mbox{\tiny EW}}^{\mbox{\tiny add}}$, together with the NLO QCD scale
uncertainty in the middle plot.}
\label{fig:nlo_qcd+ew}
\end{figure}

\begin{figure}
\centering
\begin{tabular}{cc}
\includegraphics[scale=0.45]{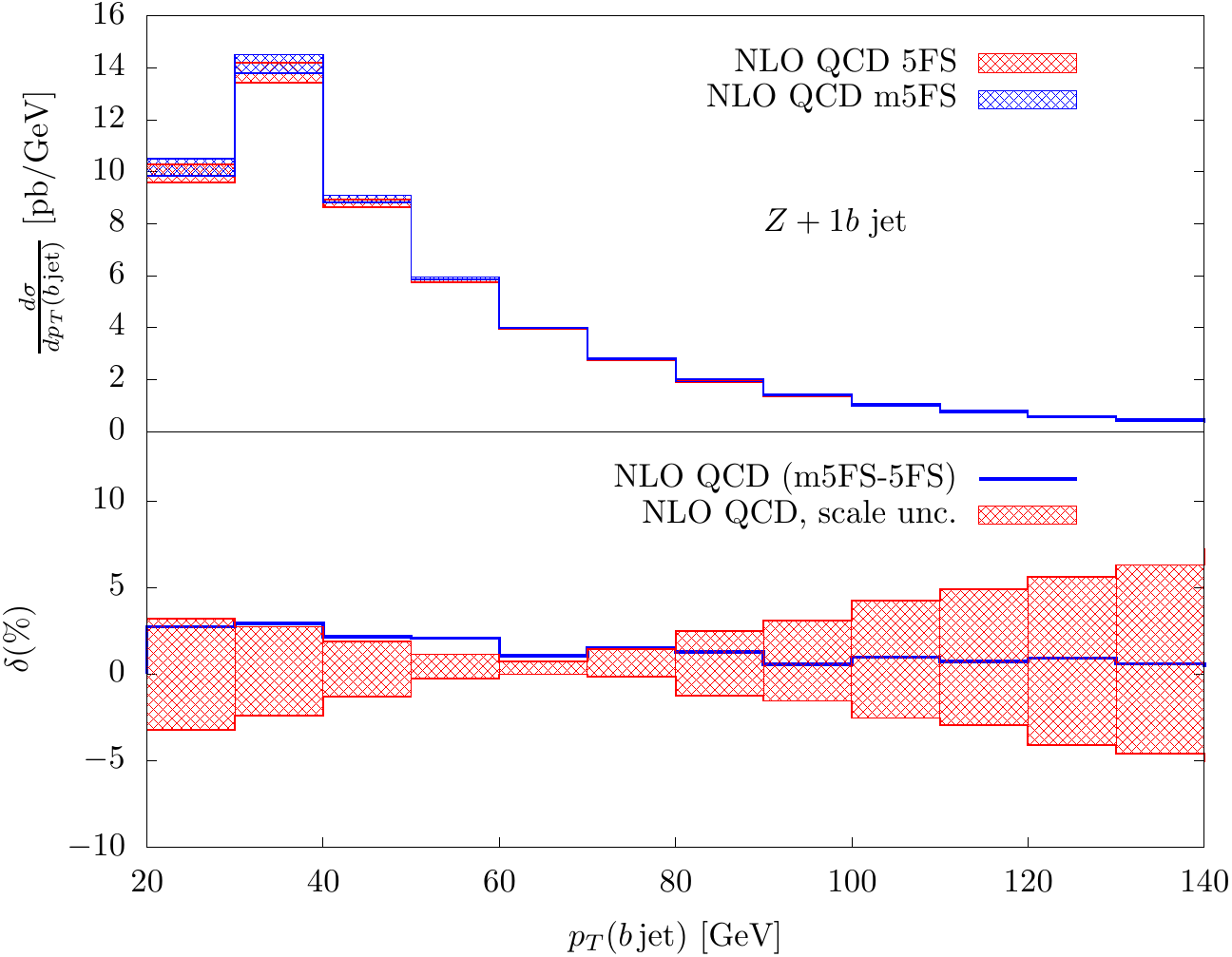} &
\includegraphics[scale=0.45]{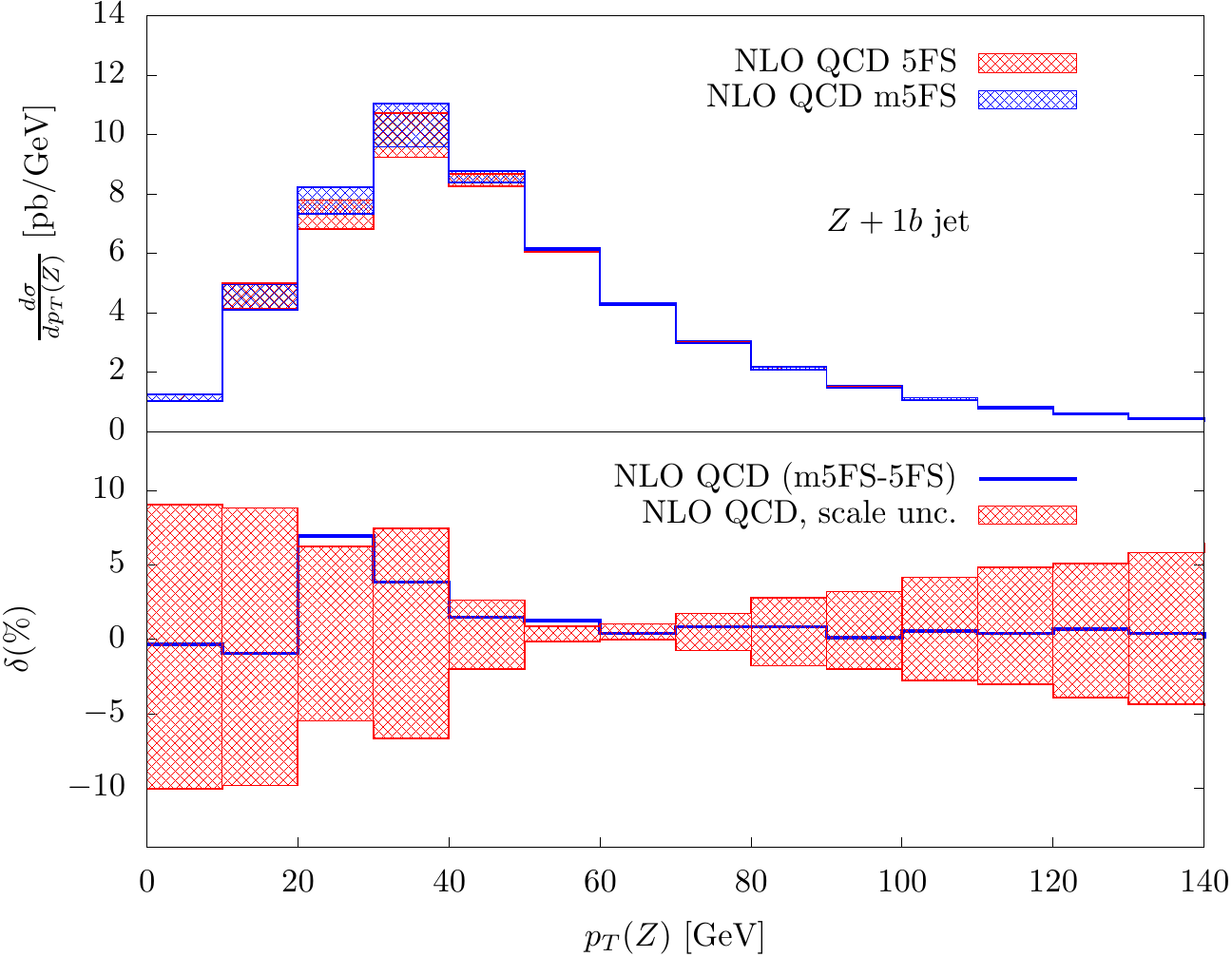}\\
\includegraphics[scale=0.45]{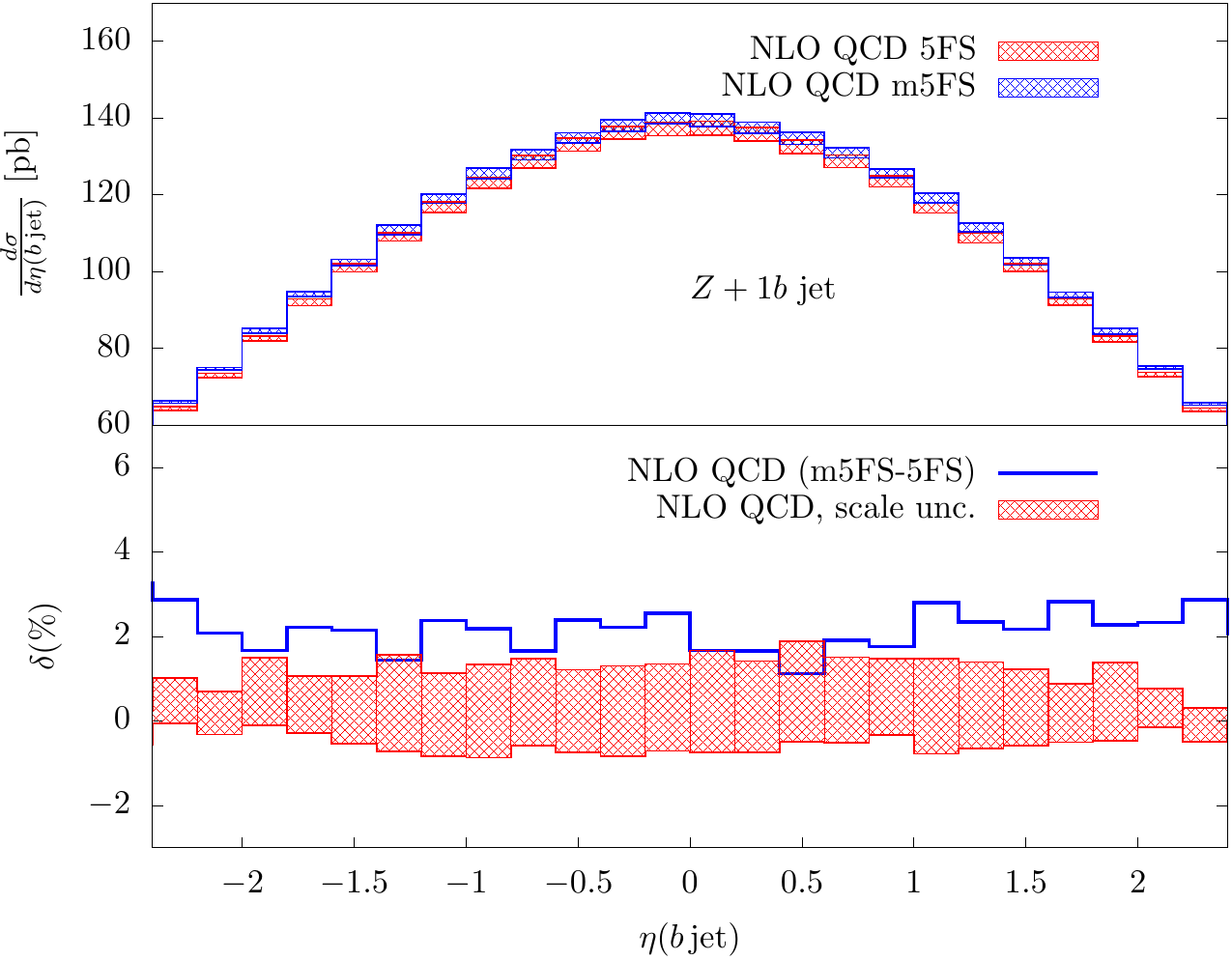} &
\includegraphics[scale=0.45]{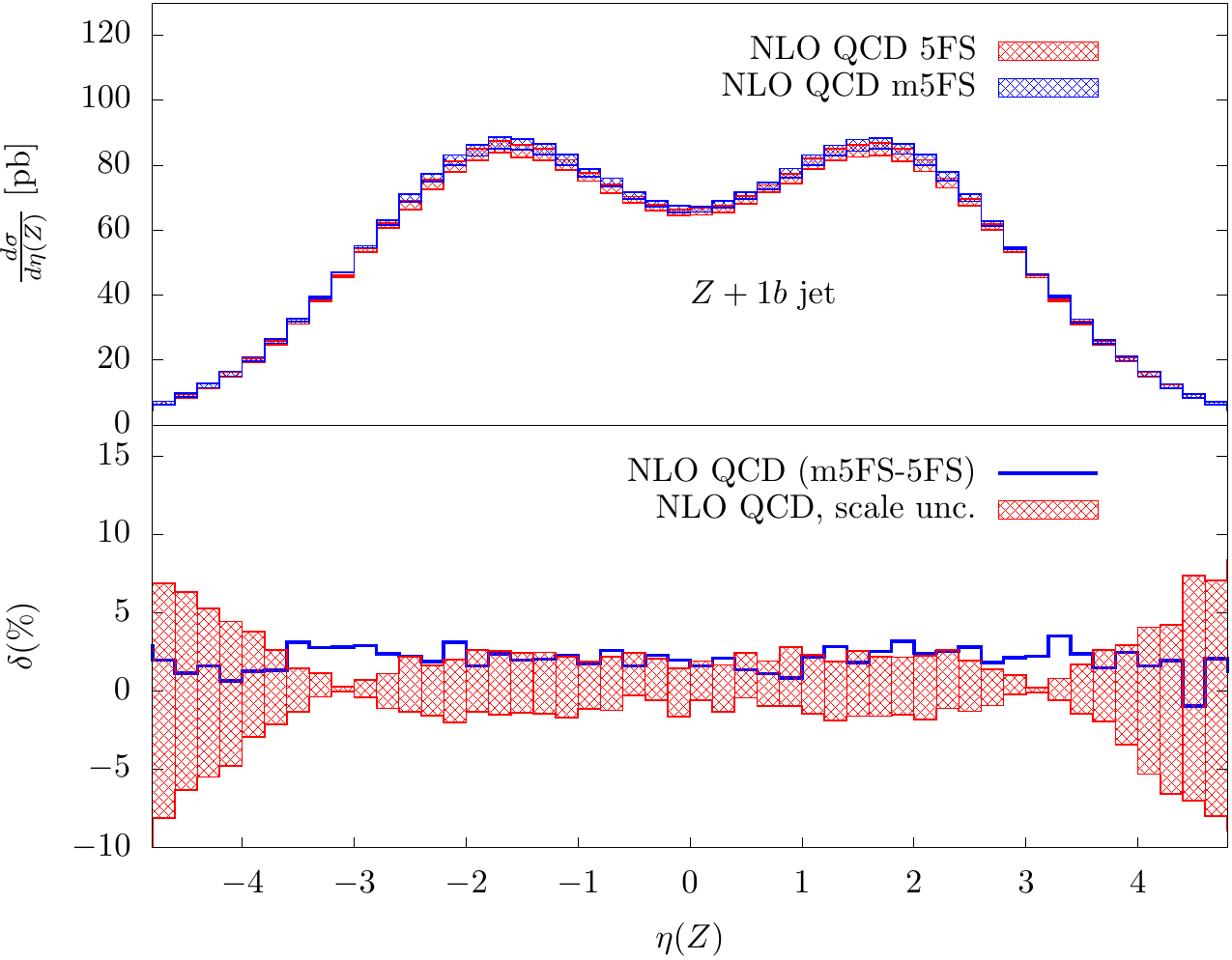}
\end{tabular}
\caption{\small 5FS and m5FS NLO QCD predictions
  for differential distributions for the $b$-jet and $Z$ transverse momentum,
  and the invariant $Z\,b$-jet mass. The lower window in each plot shows the difference
$\delta_{\mbox{\tiny mb}}$ together with the NLO QCD scale uncertainty.}
\label{fig:mb-effects}
\end{figure}

Table~\ref{tab:results} as well as Figs.~\ref{fig:nlo_qcd+ew} and
\ref{fig:mb-effects} summarize the results of
Ref.~\cite{Figueroa:2018chn} by showing both the impact of NLO EW
corrections and of initial-state $b$-quark mass effects on total and
differential cross sections. NLO QCD and EW corrections are
combined in both the \textit{additive} approach,
\begin{equation}
\sigma_{\mbox{\tiny NLO}}^{\mbox{\tiny QCD+EW}}\equiv
\sigma_{\mbox{\tiny LO}}
+\alpha_s^2\alpha\,\sigma^{(2,1)}
+\alpha_s\alpha^2\sigma^{(1,2)}\,
\end{equation}
and the \textit{multiplicative} approach
\begin{equation}
\sigma_{\mbox{\tiny NLO}}^{\mbox{\tiny QCD}\times \mbox{\tiny EW}}\equiv
\sigma_{\mbox{\tiny NLO}}^{\mbox{\tiny QCD}} \times
\frac{\sigma_{\mbox{\tiny NLO}}^{\mbox{\tiny EW}}}{\sigma_{\mbox{\tiny LO}}} 
\equiv \sigma_{\mbox{\tiny NLO}}^{\mbox{\tiny QCD}} \times
K_{\mbox{\tiny EW}}\,,
\end{equation}
where
$\sigma^{\mbox{\tiny QCD}}_{\mbox{\tiny NLO}}\equiv\sigma_{\mbox{\tiny
    LO}} + \alpha_s^2\alpha\,\sigma^{(2,1)}$
and
$\sigma^{\mbox{\tiny EW}}_{\mbox{\tiny NLO}} \equiv
\sigma_{\mbox{\tiny LO}} +\alpha_s\alpha^2\,\sigma^{(1,2)}$.
Notice that
$\sigma_{\mbox{\tiny NLO}}^{\mbox{\tiny QCD}\times \mbox{\tiny EW}}$
also includes mixed QCD$\times$EW corrections that are of higher
order. The second row of
Table~\ref{tab:results} gives the relative corrections as percentage
of the LO cross section,
$\delta(\%)=(\sigma_X/\sigma_{\mbox{\tiny LO}}-1)\times 100$ (where
X=NLO QCD, NLO EW, $\ldots$).  The quoted uncertainty is due to
varying the renormalization and factorization scales
($\mu=\mu_r=\mu_f $) between $\mu=M_Z/2$ and $\mu=2M_Z$.  The
CTEQ14qed PDF set~\cite{Schmidt:2015zda} has been used, and no PDF
uncertainty has been included in these estimates.
\begin{table}[h]
\centering
{\small
\begin{tabular}{|c|c|c|c|c|c|}
\hline
&LO & NLO QCD & NLO EW & NLO QCD+EW  & NLO QCD $\times$ EW\\ \hline \hline   
$\sigma$(pb) & 
$389.73^{+15}_{-31}(392.66^{+15}_{-32})$ & 
$537.7^{+7}_{-3}(526.9^{+9}_{-3})$ &
$383.40^{+15}_{-31}$ & 
$531.4^{+7}_{-3}$  & $529.2^{+7}_{-3}$ \\ \hline
$\delta$(\%) &
- & 38(34) & -1.6 & 36 & 36 \\\hline \hline
\end{tabular}
}
\caption{Total cross sections at LO and NLO (first row) including only QCD NLO
  corrections, only EW NLO corrections, or both, for $\mu=M_Z$ at the LHC
  with c.m. energy 13 TeV. The relative impact of each NLO
  contribution is given in the second row. The results obtained with $m_b=0$ at LO and
  NLO QCD are given in  parenthesis. See text for more details.}
\label{tab:results}
\end{table}
If the impact of NLO EW corrections on the total cross section is
indicative of the average magnitude of their effect, a much more
interesting result is their effect on distributions. Furthermore, it
is important to compare the effect of NLO EW corrections to the
residual theoretical uncertainty of NLO distributions, and estimate in
particular whether NLO EW corrections are within the scale uncertainty
of the corresponding NLO QCD corrections. This can be quantified in
terms of the following ratios:
\begin{equation}
\label{eq:delta-ew-add-mult}
\delta_{\mbox{\tiny EW}}^{\mbox{\tiny add}}=
\frac{\sigma_{\mbox{\tiny  NLO}}^{\mbox{\tiny QCD+EW}}-
\sigma_{\mbox{\tiny NLO}}^{\mbox{\tiny QCD}}}
{\sigma_{\mbox{\tiny NLO}}^{\mbox{\tiny QCD}}}
=\frac{\sigma^{\mbox{\tiny QCD+EW}}_{\mbox{\tiny NLO}}}
{\sigma^{\mbox{\tiny QCD}}_{\mbox{\tiny
      NLO}}}-1\,\,\,\,\,\,\,\mbox{and}\,\,\,\,\,\,\,
\delta_{\mbox{\tiny EW}}^{\mbox{\tiny prod}}=
\frac{\sigma_{\mbox{\tiny NLO}}^{\mbox{\tiny QCD}\times \mbox{\tiny EW}}-
\sigma_{\mbox{\tiny NLO}}^{\mbox{\tiny QCD}}}
{\sigma_{\mbox{\tiny NLO}}^{\mbox{\tiny QCD}}}
=\frac{\sigma^{\mbox{\tiny EW}}_{\mbox{\tiny NLO}}}
{\sigma_{\mbox{\tiny LO}}}-1\,,
\end{equation}
for to the \textit{additive} and \textit{multiplicative}
approach of combining NLO QCD and EW corrections respectively.
Following the same logic, we quantify the effect of switching from a
massless 5FS to a m5FS in terms of the ratio
\begin{equation}
\delta_{\mbox{\tiny mb}}=
\frac{\sigma_{\mbox{\tiny NLO}}^{\mbox{\tiny QCD,m5FS}}-
\sigma_{\mbox{\tiny NLO}}^{\mbox{\tiny QCD,5FS}}}
{\sigma_{\mbox{\tiny NLO}}^{\mbox{\tiny QCD,5FS}}}
=\frac{\sigma_{\mbox{\tiny NLO}}^{\mbox{\tiny QCD,m5FS}}}
{\sigma_{\mbox{\tiny NLO}}^{\mbox{\tiny QCD,5FS}}}-1\,,
\end{equation}
representing the fractional change in the NLO QCD cross sections in
going from the 5FS to the m5FS.
It is clear that both mass effects and NLO EW corrections are small
effects, compared to the size of NLO QCD corrections, and are mostly
within the uncertainty of the NLO QCD cross section.  Hence the need
for a better control of higher-order QCD corrections. Still, there are
clear indications of the impact of EW corrections and mass effects,
in particular on angular distributions, and in complementary kinematic
regions: EW corrections mainly affect the high-$p_T$ region of both
$b$-jet and $Z$ boson $p_T$, while $b$-quark mass effects are more
visible in the low-$p_T$ regions. Interestingly, they both tend to be
more visible in angular distributions, such as pseudorapidity
distributions.  One would indeed expect that a modification of the
initial-state kinematics can lead to modifications of the final-state
angular distributions. This corroborates the original motivation for a
consistent implementation of a m5FS in Monte Carlo event generators.

\bibliographystyle{JHEP}
\bibliography{ll2018-reina}

\end{document}